\documentclass{eas}
\usepackage{graphicx}
\begin{document}

\title{The nearby AGB star L$_2$ Puppis: the birth of a planetary nebula ?} 
\runningtitle{NACO and SPHERE observations of L$_2$ Puppis}
\author{P.~Kervella}\address{Unidad Mixta Internacional Franco-Chilena de Astronom\'{i}a (UMI 3386), CNRS/INSU, France \& Departamento de Astronom\'{i}a, Universidad de Chile, Camino El Observatorio 1515, Las Condes, Santiago, Chile.}
\secondaddress{LESIA (UMR 8109), Observatoire de Paris, PSL, CNRS, UPMC, Univ. Paris-Diderot, 5 place Jules Janssen, 92195 Meudon, France}
\author{M.~Montarg\`es}\address{Institut de Radioastronomie Millim\'etrique, 300 rue de la Piscine, 38406, Saint Martin d'H\`eres, France}\sameaddress{, 2}
\author{E.~Lagadec}\address{Laboratoire Lagrange, Universit\'e C\^ote d'Azur, Observatoire de la C\^ote d'Azur, CNRS, Bd de l'Observatoire, CS 34229, 06304 Nice Cedex 4, France}
\begin{abstract}
Adaptive optics observations in the infrared (VLT/NACO, Kervella et al. 2014) and visible (VLT/SPHERE, Kervella et al. 2015) domains revealed that the nearby AGB star L$_2$ Pup ($d=64$\,pc) is surrounded by a dust disk seen almost edge-on. Thermal emission from a large dust "loop" is detected at 4\,$\mu$m up to more than 10\,AU from the star. We also detect a secondary source at a separation of 32\,mas, whose nature is uncertain. L$_2$ Pup is currently a relatively "young" AGB star, so we may witness the formation of a planetary nebula. The mechanism that breaks the spherical symmetry of mass loss is currently uncertain, but we propose that the dust disk and companion are key elements in the shaping of the bipolar structure. L$_2$ Pup emerges as an important system to test this hypothesis.
\end{abstract}
\maketitle
\section{Introduction}
The final stages of the evolution of low and intermediate mass stars involve complex and poorly understood physical phenomena that lead to the formation of axially symmetric bipolar planetary nebulae (PNe). The origin of these spectacular structures is generally associated to interactions between the material expelled from the evolved star and an orbiting companion (Lagagec \& Chesneau \cite{lagadec14}; Soker \& Livio \cite{soker89}). The presence of a circumstellar disk is also invoked to collimate mass loss and foster the appearance of the axial symmetry, but such disks have only been observed in a few evolved stars.
L$_2$ Pup (HD\,56096) is one of the nearest ($d=64\,pc$, van Leeuwen \cite{vanleeuwen07}) and brightest ($m_V \approx 7$, $m_K \approx -1.5$) asymptotic giant branch (AGB) star that exhibits Mira-like pulsations with a period of 141\,days (Bedding et al.~\cite{bedding02}).
This star is particularly interesting as Kervella et al.~(\cite{kervella14}) discovered a circumstellar disk around L$_2$\,Pup using adaptive optics in the near-infrared. This discovery was recently confirmed from polarimetric imaging at visible wavelengths by Kervella et al.~(\cite{kervella15}), that also revealed the presence of a secondary source. We present these two sets of observations in Sect.~\ref{observations}, and we discuss the spectral energy distribution and disk geometry in Sect.~\ref{sed}.

\section{Adaptive optics observations \label{observations}}
\subsection{Infrared VLT/NACO imaging}
\begin{figure}[htbp]
\begin{center}
\frame{\includegraphics[width=4cm]{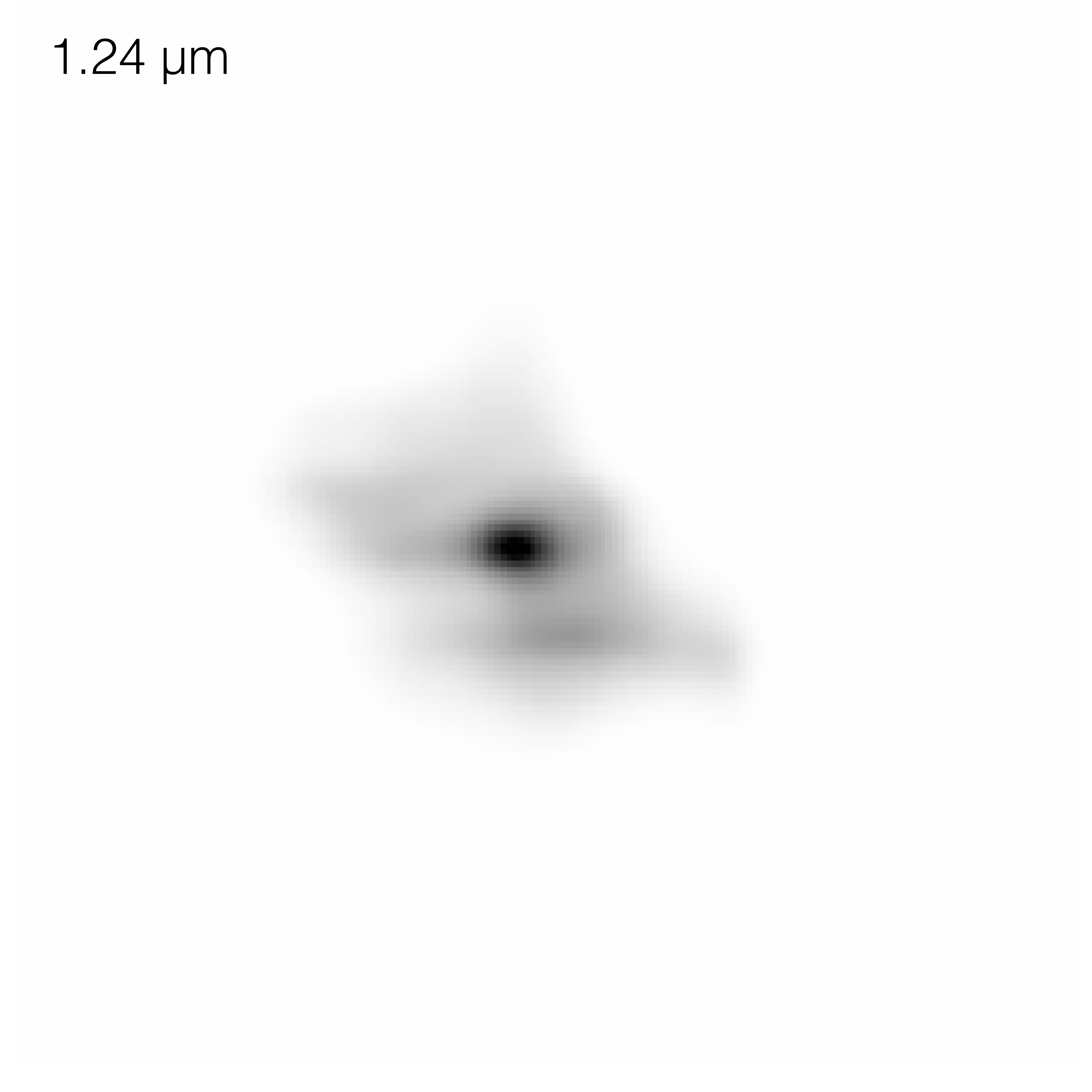}}
\frame{\includegraphics[width=4cm]{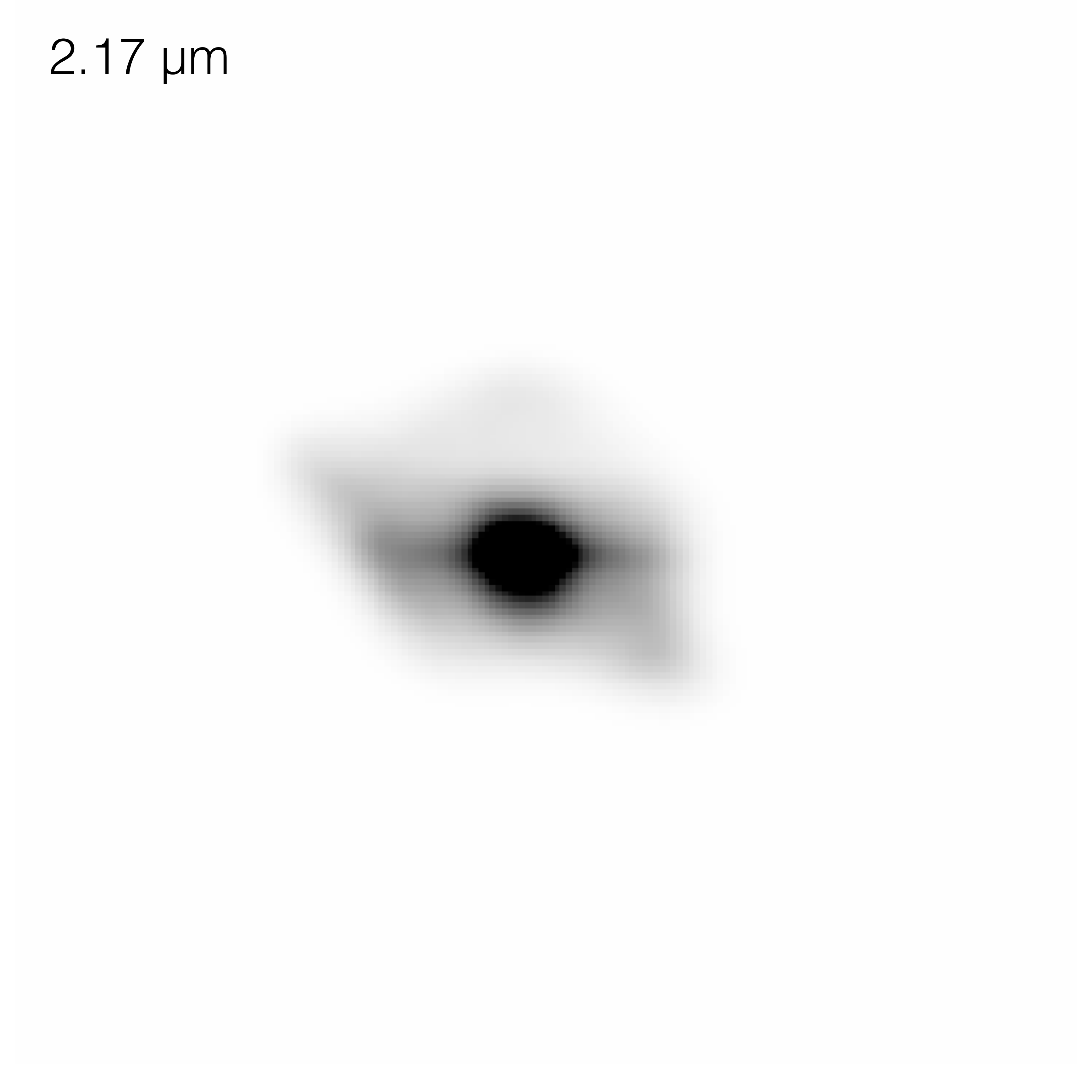}}
\frame{\includegraphics[width=4cm]{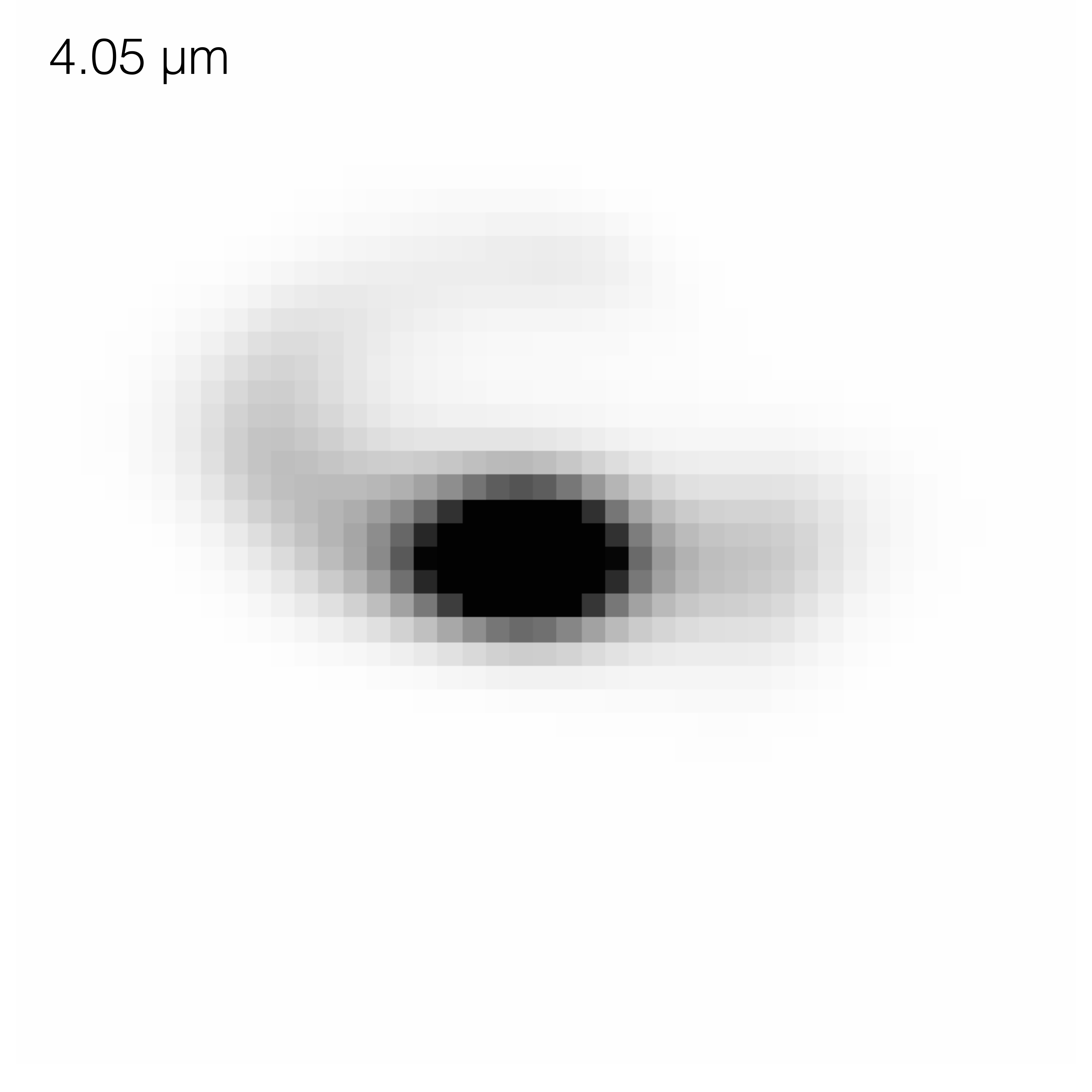}}
\caption{Aspect of L$_2$\,Pup in the infrared (field of view 0.6'', North up and East to the left, from Kervella et al.~\cite{kervella14}).}
\label{naco}
\end{center}
\end{figure}

Kervella et al.~(\cite{kervella14}) reported observations of L$_2$ Pup using the VLT/NACO (Rousset et al.~\cite{rousset03}) adaptive optics system. Using a ``lucky imaging'' approach, they obtained diffraction limited images in a series of narrow-band filters ranging in wavelength from 1 to 4\,$\mu$m. These images show the presence of a dust band in front of L$_2$ Pup. The band exhibits a high opacity in the $J$ band ($\lambda \approx 1\,\mu$m) and becomes translucent in the $H$ and $K$ bands ($1.6-2.2\,\mu$m) where the light scattering becomes less efficient (Fig.~\ref{naco}). In the $L$ band, the thermal emission from the inner rim of the dust disk was also detected, as well as a large loop extending to more than 10\,AU from the star.

New observations by Lykou et al.~(\cite{lykou15}) using the sparse aperture masking mode of NACO also showed the presence of a disk-like structure, and Ohnaka et al.~(\cite{ohnaka15}) confirmed the presence of an elongated central emission in the east-west direction. 

\subsection{Visible VLT/SPHERE polarimetric imaging}
\begin{figure}[htbp]
\begin{center}
\includegraphics[width=4cm]{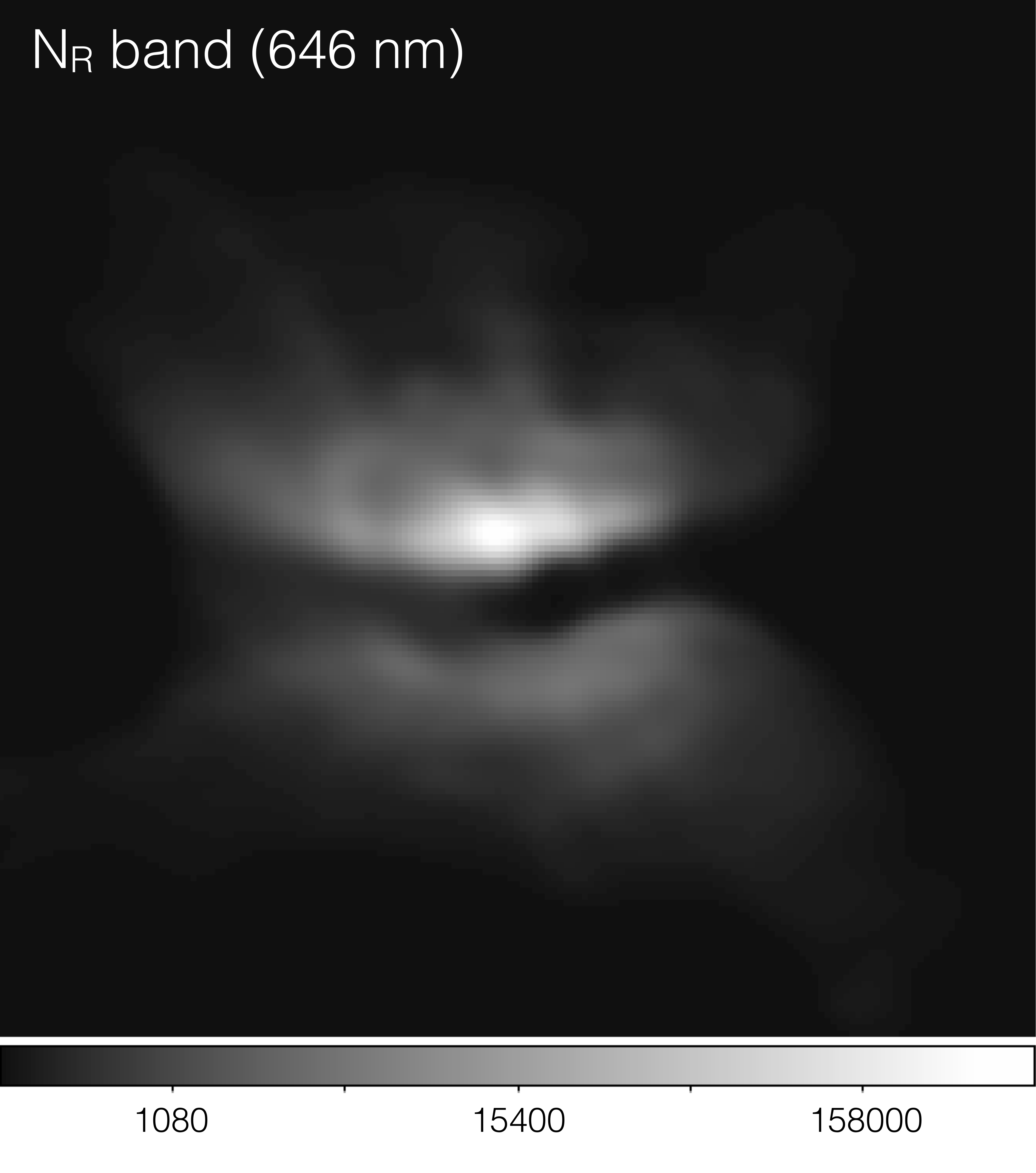}
\includegraphics[width=4cm]{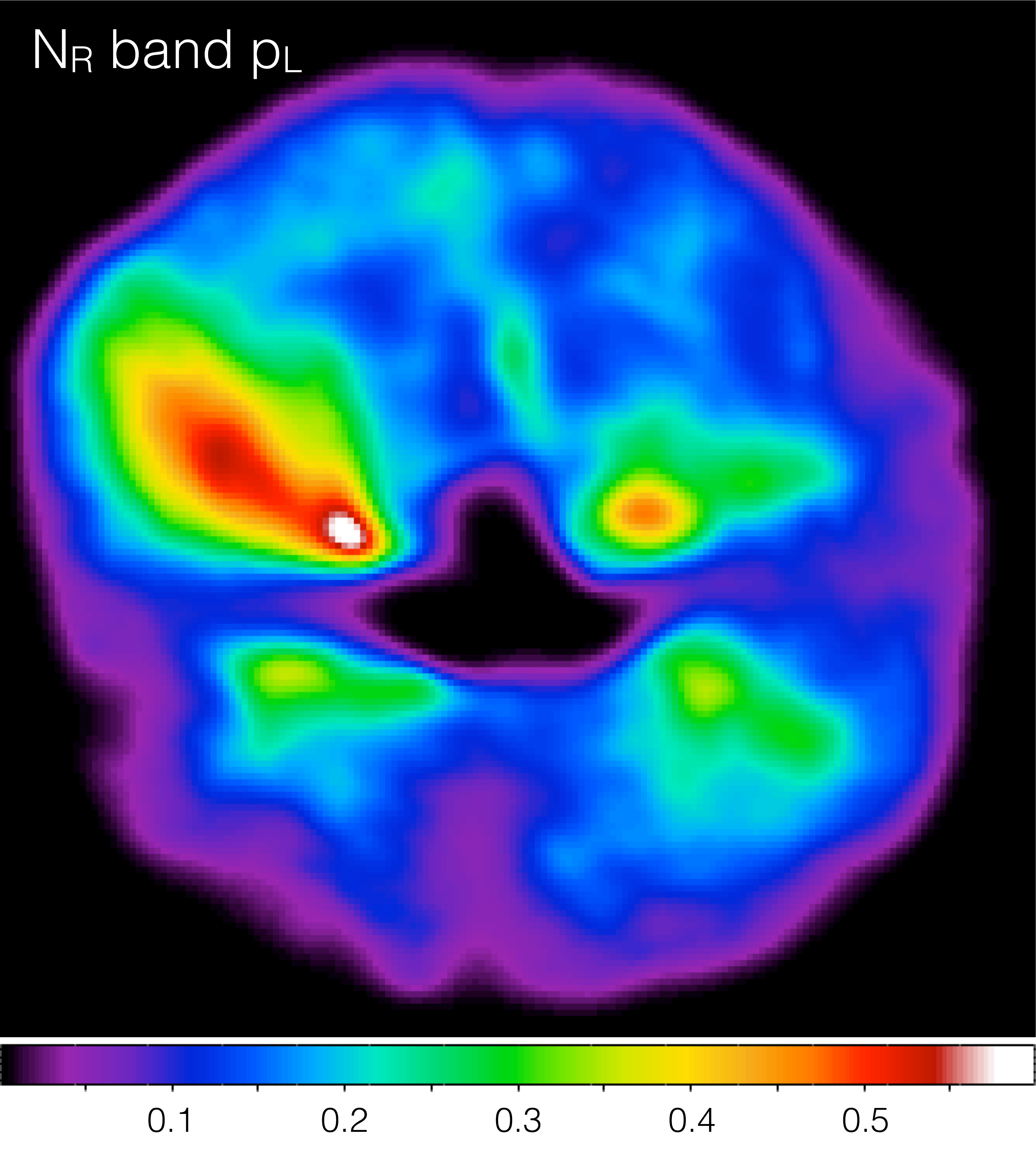}
\includegraphics[width=4cm]{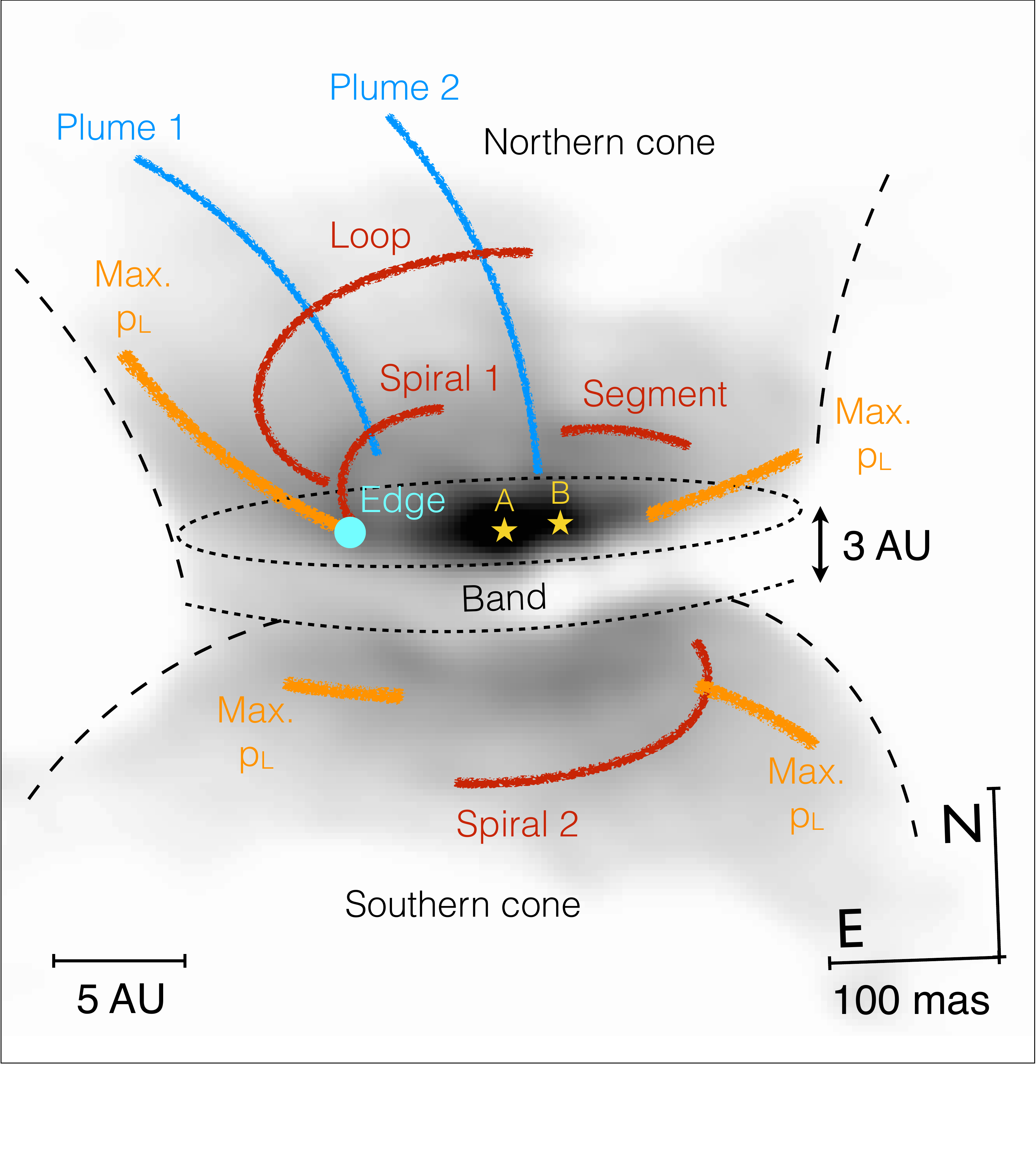}
\caption{Intensity (ADU, left) and degree of linear polarization (middle) maps of L$_2$\,Pup. The right panel shows the observed features (field of view 0.6'', Kervella et al.~\cite{kervella15}).}
\label{zimpol}
\end{center}
\end{figure}

The Spectro-Polarimetric High-contrast Exoplanet REsearch (SPHERE, Beuzit et al.~\cite{beuzit08}) is a high performance adaptive optics (Fusco et al.~\cite{fusco14}) recently installed at the Very Large Telescope. Kervella et al.~(\cite{kervella15}) observed L$_2$ Pup using SPHERE equipped with the imaging polarimeter ZIMPOL (Roelfsema et al.~\cite{roelfsema14}). This instrument provides diffraction limited imaging at visible wavelengths, down to $\lambda \approx 550$\,nm. The observed intensity image and the map of the degree of linear polarization in the $N_R$ band ($\lambda = 646$\,nm) are presented in Fig.~\ref{zimpol}, together with a nomenclature of the observed features.

The intensity image shows a bright unresolved source at a separation of 0.032'' from the AGB star (labeled 'B' in Fig.~\ref{zimpol}, right panel). Although the nature of this source is unknown, its brightness and color are consistent with a late K giant star. Several spiral structures are also observed, as well as two thin plumes extending perpendicularly to the disk plane. The map of the degree of linear polarization $p_L$ shows a maximum at a radius of 6\,AU from the central star. Kervella et al.~(\cite{kervella15}) interpret this maximum as 90$^\circ$ scattering occurring at the inner rim of the dust disk. The radius is consistent with the observed extension of the thermal emission in the $K$ band (Fig.~\ref{naco}, central panel).

\section{Spectral energy distribution and modeling \label{sed}}

\begin{figure}[htbp]
\begin{center}
\includegraphics[height=4.2cm]{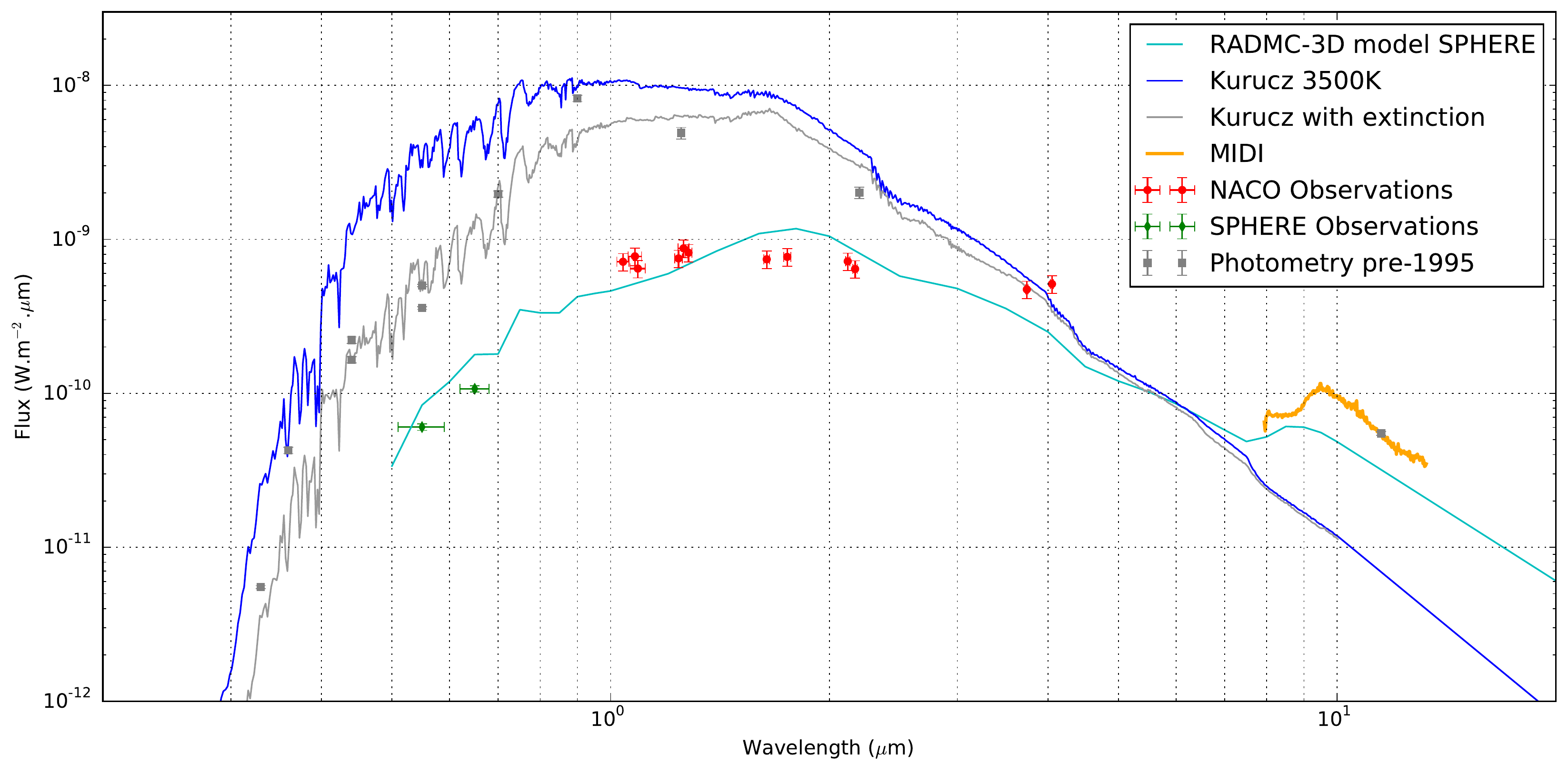}
\includegraphics[height=4.2cm]{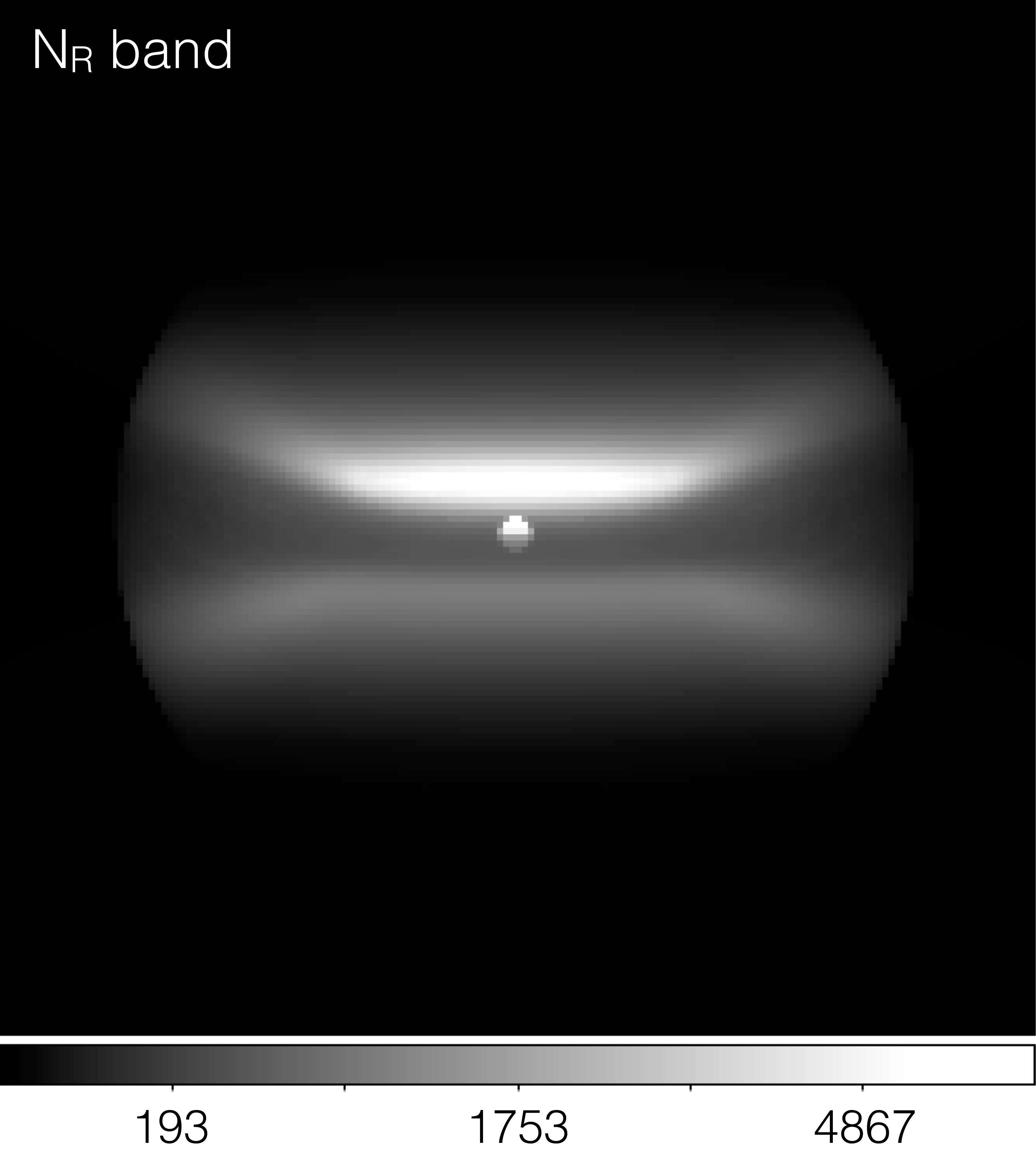}
\caption{Spectral energy distribution (left) and RADMC-3D model of L$_2$ Pup's disk (right, taken from Kervella et al.~\cite{kervella15}).}
\label{sedmodel}
\end{center}
\end{figure}

The radiative transfer modeling (using the RADMC-3D code; Dullemond \cite{dullemond12}) presented by Kervella et al.~(\cite{kervella15}, Fig.~\ref{sedmodel}, right panel) shows that the circumstellar disk is seen almost edge on, with an inclination of $\approx 82^\circ$ on the line of sight. This model reproduces satisfactorily the observed spectral energy distribution (Fig.~\ref{sedmodel}, left), that exhibits a remarkably flat flux density between 1 and $4\,\mu$m. The total dust mass in the RADMC-3D disk model is $2.4\ 10^{-7}$~M$_\odot$. Assuming a gas-to-dust ratio of 100, this translates to a total mass including gas of $10^{-5}$ to $10^{-4}$~M$_\odot$.

\section{Conclusion}
The environment of L$_2$ Pup hosts the two major ingredients suspected to play a role in the shaping of PNe: a close-in companion (whose physical nature is still largely enigmatic) and a circumstellar dust disk. Thanks to its proximity and the availability of the SPHERE adaptive optics providing 20\,mas angular resolution in the visible, it is now possible to monitor in real time the dynamical evolution of the different components of the immediate environment of L$_2$ Pup (companion, spirals, plumes,...). Considering their linear scales, we expect a significant evolution over timescales of only a few years. L$_2$ Pup thus exhibits a strong potential to improve our understanding of the mass loss of low and intermediate mass stars and the formation of bipolar PNe.

\section{Questions}

\noindent
{\it J.-Ph. Berger:} If you have such a massive companion you should have strong tidal truncation and a cavity. It is surprising not to see any sign of cavity in the spectral energy distribution (SED).\\
{\it Answer:} The SED of L$_2$ Pup is atypical due to its flatness over a major part of the infrared domain ($1-4\,\mu$m). The fact that the disk is seen almost edge-on affects the flux contribution from the inner rim of the disk through absorption and reddening, as shown by the RADMC-3D model. This may explain that the signature of the cavity is not immediately observable in the SED.

\medskip
\noindent
{\it J. Groh:} Is this object representative of its class, or is it a peculiar object ? Would it be possible to observe a larger sample to investigate this phenomenology ?\\
{\it Answer:} The AGB star itself appears to be close in physical properties to the Mira star R Vir, whose pulsation period is 145.5 days (see e.g. Eisner et al.~\cite{eisner07}). More generally, L$_2$ Pup A could well be classified as a low amplitude Mira (Kervella et al.~\cite{kervella14}). The presence of circumstellar disk has been proposed around the Mira V CVn by Neilson et al.~(\cite{neilson14}) based on polarimetry, but this is a rare example of such a detection around a Mira. The difficulty to extend the sample is the angular scale of the disks: L$_2$ Pup is very nearby, so the angular scale is resolvable using SPHERE, but for more distant stars, optical or millimeter interferometry would be the only options.

\medskip
\noindent
{\it A. Lobel:} When you overplot the $4\,\mu$m image showing the extended loop, it appears to be projected on the Northern cone. Could it tell about the direction of the orbital motion on the sky ?\\
{\it Answer:} There is unfortunately an ambiguity whether the loop is located on the nearby or far surface of the Northern cone. The difficulty is that the thermal emission is isotropic, while the scattering is preferentially in the forward direction. We therefore very likely observe in Fig.~\ref{zimpol} the nearby side of the cones, but we cannot be sure for the thermal emission of the loop.

\medskip
\noindent
{\it T. Ueta:} 1) Bright spots A and B in the ZIMPOL image could still be just brightly lit surfaces of the top of the disk. Polarization vector maps should help pinpoint the location of the central bright source. Does the centrosymmetric structure of vector maps point to the location of A ?\\
2) The distribution of scattering matter can be better probed by the polarized flux $I_\mathrm{pol}$ rather than the degree of polarization $p_L$. The disk edge might show differently in $I_\mathrm{pol}$ than in $p_L$, or do they look the same ?\\
{\it Answer:} 1) The nature of the secondary source is indeed uncertain. But its flux contribution is considerable in the visible ($\approx 20$\% of the primary), pointing at a stellar origin. Regarding the polarization vector, the difficulty is that the secondary is located so close to the primary (32 mas), that its measurement is highly uncertain. We see a slight deviation from the classical single central source vector map (purely orthoradial), but not at a sufficiently significant level.\\
2) The degree of polarization $p_L$ has the advantage to be a simple proxy for the scattering angle $\theta$ (in the single scattering approximation), and therefore well suited to probe the overall geometry of the disk. The polarized flux $I_\mathrm{pol}$ is sensitive to both the scattering angle $\theta$ and the scattering dust density $\rho$, therefore relatively more complex to model. In L$_2$ Pup, the $I_\mathrm{pol}$ map is generally similar to the total intensity map, except close to the primary star where the polarization is low (forward scattering). The secondary source exhibits a relatively strong $I_\mathrm{pol}$ contribution, suggesting that it contains a significant fraction of scattered light from the primary. But its overall degree of polarization $p_L$ is low, indicating that it is not purely scattering. The presence of an accreting disk around a companion star is a possibility that we plan to investigate.

\medskip
\noindent
{\it J. Milli:} Is the angle of polarization deduced from ZIMPOL observations consistent with the morphological model ?\\
{\it Answer:} For most of the field of view apart from the very central part, the polarization vector is orthoradial, consistent with a central source illuminating the dust envelope (as in the RADMC-3D model). But very close to the star (within $\approx 50$\,mas), the determination of the polarization angle is uncertain (see also the answer to Dr.~Ueta's question 1).

\medskip
\noindent
{\it J. Hoffman:} Does your radiative transfer model predict your observed polarization results ?\\
{\it Answer:} We have encountered some difficulties with the numerical computation of the polarization with the RADMC-3D code. But we indeed plan to compare the observed polarization maps to the predictions of the model.


\end{document}